\documentclass{aa}
\usepackage{graphicx} 
\begin{document}

\title{On the generation of asymmetric warps in disk galaxies\\}
\author{Kanak Saha \&  Chanda J. Jog}

\offprints{K. Saha}
\institute{Department of Physics, Indian Institute of Science, Bangalore 560012, 
India\\
\email{kanak@physics.iisc.ernet.in,cjjog@physics.iisc.ernet.in}}
\date{Received; accepted}

\abstract
{The warps in many spiral galaxies are now known to asymmetric. Recent sensitive observations have revealed that asymmetry of warps may be the norm rather than exception. However there exists no generic mechanism to generate these asymmetries in warps.}
{To provide an explanation for the generation of asymmetric warps in disk galaxies}
{We have derived the dispersion relation in a compact form for the S-shaped warps(described by the $m=1$ mode) and the bowl-shaped distribution(described by the $m=0$ mode) in galactic disk embedded in a dark matter halo. We then performed the numerical modal analysis and used the linear and time-dependent superposition principle to generate asymmetric warps in the disk.}
{On doing the modal analysis we find the frequency of the $m=0$ mode is much larger than that of the $m=1$ mode. The linear and time-dependent superposition of these modes with their unmodulated amplitudes(that is, the coefficients of superposition being unity) results in an asymmetry in warps of $\sim 20 - 40 \%$, whereas a smaller coefficient for the $m=0$ mode results in a smaller asymmetry. The resulting values agree well with the recent observations. We study the dependence of the asymmetry index on the dark matter halo parameters. This approach can also naturally produce U-shaped warps and L-shaped warps.}
{We show that a rich variety of possible asymmetries in the z-distribution of the spiral galaxies can naturally arise due to a dynamical wave interference between the first two bending modes(i.e. $m=0$ and $m=1$) in the disk. This is a simple but general method for generating asymmetric warps that is independent of how the individual modes arise in the disk.} 
\keywords{galaxies: kinematics and dynamics - galaxies: spiral - galaxies: structure}

\titlerunning {On the generation of asymmetric warps}

\maketitle

\section{Introduction}
The disks in most spiral galaxies are not flat and their outer parts often 
show warping of the material away from the galactic midplane. Warps are 
mostly seen in the 
atomic hydrogen gas (HI) in the outer parts of a galactic disk (Sancisi 1976, Bosma 1978), but there is also a strong observational evidence for 
optical warps in old stars, as in NGC 4565 (vand der Kruit \& Searle 1981). 
A statistical analysis of a large number of edge-on spirals has shown 
that optical warps are seen in a significant fraction (about 40 \%) of the spiral galaxy disks 
(Sanchez-Saavedra et al. 1990, Reshetnikov \& Combes 1998, 1999). 

Thus the warps are a common phenomenon and hence must be either long-lived 
or be excited repeatedly. Tidal interactions have been proposed as a
triggering mechanism for warps, however since many isolated spirals- such as 
NGC 4565 also show warps (Sancisi 1976), this cannot be the main triggering 
mechanism for warps. Anisotropic pressure applied by the intergalactic magnetic field can generate an S-shaped warp in the galactic disk (Battaner et al. 1990). But a very high value of the magnetic field is required to produce warp near 4-6 disk scalelengths. Alternatively gas accretion could generate warps 
repeatedly, as for example seen in the cosmological simulations of disk 
formation (Semelin \& Combes 2005).
 
One plausible mechanism to get long-lived warps
is to use the modal approach first proposed and studied by 
Hunter \& Toomre (1969). Sparke \& Casertano (1988) (hereafter, SC88) followed Hunter \& Toomre's (1969) approach but also included a dark matter halo
and achieved the above goal of avoiding differential precession.
They have shown that in the presence of an oblate halo potential
a system of concentric circular rings, representing the galactic
disk, is able to accomplish a configuration where all the rings
are synchronized to precess with a uniform precession frequency.
Such modes are called a normal modes, and in this case since the
halo is included,  it is a
modified-tilt mode of the disk because it is derived by
modifying the trivial tilt mode solution of the linearized
equations of motion of an isolated system of rings. 

The warps were initially deduced to have an S-shaped distribution, with
equal amplitudes on both sides of the mid-plane. This makes their theoretical analysis easier. 
It has long been known, however, that quite a few galaxies show asymmetric 
warps so that the amplitudes on the two sides are appreciably different.
For example, our own Galaxy has an asymmetric warp: the mean plane 
bends up to 4 kpc above the plane of the inner Galaxy on the northern 
side of the disk, while it reaches down only half on the southern side and 
then turns back up toward the inner disk plane (see Burton 1988).
 More recent, sensitive observations have revealed that asymmetry
of warps may be the norm rather than the exception as shown by 
the HI gas and optical R-band observations (Garcia-Ruiz
et al. 2002). A detailed catalog of the measure of asymmetries in optical warps 
seen in edge-on galaxies can be found in Sanchez-Saavedra et al. (2003). 
Theoretically, S-shaped as well as asymmetric warps have been
shown to arise via bending instabilities  in the N-body studies of
Revaz \& Pfenniger (2004); also these have been proposed
 to arise due to gas infall (Lopez-Corredoira et al. 2002).

In this paper we propose and study the generation of the asymmetric
warps in disk galaxies. We propose that an asymmetric warp arises due to a
dynamical wave interference between the first two stable 
bending modes(i.e. $m=0$ and $m=1$) in a gravitating 
system. We do not deal with the orign of these modes, we assume
that some excitation mechanism - perhaps an external perturber, extragalactic magnetic field, or gas accretion,  or internal bending instabilities -
gives rise to both these modes and hence we do not include explicitly any source term in our analysis. It is reasonable that, once excited, these modes will interact with each other. We study a general case  so that
the modes need not even be triggered in a commensurate fashion. 
A different epoch of onset for the two modes for the two cases will merely be
reflected in a different initial ratio of the amplitudes of the modes.

It was demonstrated numerically by SC88 and Sparke (1995) respectively that
warping mode ($m=1$) and the bowl-shaped mode ($m=0$) are stable in a cold, 
thin self-gravitatiing axisymmetric disk. The basic idea of the present
paper is to use this fact 
and then let these modes interfere with each other, and see if this results
in an asymmetric warp.
This idea of superposition of modes to explain the origin of
asymmetric warps was first proposed by Sparke (1995), and also 
more recently  by Lopez-Corredoira et al (2002) and
Castro-Rodriguez et al. (2002), 
but they did not work out further details.

We do not address here the other important aspect of the 
warping phenomenon, namely its maintenance, which is a long-standing problem. 
 In the modal approach some reasons for the reduction of warp lifetime are:
back-reaction of the uniformly precessing disk 
onto a live dark matter halo (Nelson \& Tremaine 1995), and 
disk thickness and random motion (Sellwood 1996),
which can damp out the disk warping in a few orbital time scales.
Some numerical simulation studies (Dubinsky \& Kuijken 1995; 
Binney et al. 1998) have confirmed  that the dynamical 
friction due to the oblate dark matter halo  damps out a symmetric warp. On the other hand, the maintenance problem does not arise if there is gas accretion of $\sim$ few $M_{\odot} yr^{-1}$ repeatedly generating warps as discussed above. 

In this paper, we have isolated and studied one well-defined physical aspect
of the problem, namely that if the m=1 and m=0 modes are excited due to 
any of the mechanisms mentioned above- namely, tidal
interaction, or  gas accretion, or bending instabilities,  then it is inevitable
that these will get superposed and we study the resulting behavior.
Further, even if these modes are short-lived, if these can be excited 
repeatedly, then we can explain why a large fraction of spiral galaxies 
show asymmetric warps.

In section 2, we formulate the equations for $m=0$ and $m=1$ cases, 
and give the numerical scheme for solving these, and also specify the
input parameters for the disk and the halo. The results are
described in Section 3, and Section 4 summarizes the conclusions
from this paper.

\section {\bf {Formulation:} }
            
\subsection {\bf {Dynamics of bending modes:} }

We consider various possible bending modes in a 
cold, self-gravitating, thin axisymmetric 
disk in the presence of a  rigid or non-responsive dark matter halo. 
The disk,with a radial surface density profile $\Sigma(r)$, 
rotates in the equatorial plane ($z=0$) of the spheroidal dark matter 
halo with angular speed $\Omega(r)$ about the halo's symmetry
axis ($r=0$). 
Here $(r,\varphi,z)$ are the circular cylindrical polar coordinates.
The dynamical system described above is the same as in SC88 
and most of the formulation follows SC88.

The dynamical equation of a small bending of the disk perpendicular to 
its unperturbed plane($z=0$) is given by

$${ \left( {\frac{\partial}{\partial t}} + \Omega(r){\frac{\partial}{\partial \varphi}}\right)}^2{\mathcal Z} \:=\: {{\mathcal F}_{self}} +{{\mathcal F}_{halo}} \eqno(1) $$  

\noindent where the small bending is described by a single function ${\mathcal Z}(r,\varphi,t)$ and ${\mathcal F}_{self}$ is the vertical force due to the bent disk itself.
${{\mathcal F}_{halo}}$ is the vertical restoring force near the
disk plane ($z=0$) due to the dark matter halo.

We consider the small bending of the disk normal to its plane as resulting 
from the linear superposition of different bending modes each described by 
an azimuthal wave number $m$. So the linear superposition allows us to write 

$$ {{\mathcal Z}(r,\varphi,t)} \: = \: {\sum_{m}}{A_{m} z_{m}(r,\varphi,t)} \eqno(2a) $$

In the above relation $A_{m}$ is a constant number which denotes the coefficients of superposition and 
$$z_{m}(r,\varphi,t)\: = 
\: \Re\{ h_{m}(r) e^{i(\omega_{m} t - m\varphi )}\} \eqno(2b) $$ 
describes the bending mode of the disk with azimuthal wave
number $m$, where $h_{m}(r)$ is the unmodulated amplitude of the
m$^{th}$ mode. Thus, the total amplitude of the $m^{th}$ mode in the superposition is given by $A_m \: h_m (r) $.

The general dynamical equation for the bending mode with the azimuthal 
wavenumber $m$ is then obtained by substituting the form given
by eq. (2b) into eq. (1) :
\begin{eqnarray}
\lefteqn{ \left[ {\left(\omega_{m} - m\Omega(r)\right)}^2 - {\nu_{h}}^2 (r) \right]h_{m}(r) = } \nonumber \\
& & \: \: \: \: \: \: \: \: \: \: \: \: \: \: \: \: \: \: \: \: \: \: \: \: \: \: G{\int_{0}^{\infty}}{ \Sigma(r^{\prime})H(r,r^{\prime})[h_{m}(r) -h_{m}(r^{\prime})]r^{\prime} dr^{\prime} } \nonumber \\
& & \: \: \: \: \: \: \: \: \: \: \: \: \: \: \: \: \: \: \: \: \: \: \: \: \: \: \: \: \: \: +G{\int_{0}^{\infty}} { \Sigma(r^{\prime}) K_{m}(r,r^{\prime}) h_{m}(r^\prime) r^{\prime} dr^{\prime} } \nonumber \: \: \: (3a)
\end{eqnarray}
The angular speed $\Omega(r)$ appearing in above equation gets a contribution from the warped disk as well as from the dark matter halo described in sec. 2.3.
The definitions of $H(r,r^{\prime})$ and $K_{m}(r,r^{\prime})$ are as follows:
$$ H(r,r^{\prime}) \:=\: \int_{0}^{\infty}{\frac{1}{[r^2 +{r^{\prime}}^2 -2rr^{\prime}{\cos{\psi}} + {z_{0}}^2]^{\frac{3}{2}}}} d{\psi} \eqno(3b)$$

$$ K_{m}(r,r^{\prime}) \:=\: \int_{0}^{\infty}{\frac{1-\cos(m\psi)}{[r^2 +{r^{\prime}}^2 -2rr^{\prime}{\cos{\psi}} + {z_{0}}^2]^{\frac{3}{2}}}} d{\psi}  \eqno(3c) $$
These functions can be further expressed in terms of complete elliptical integrals.
Note that for the bowl-shaped mode($m=0$), the function  $K_{0}(r,r^{\prime})$ vanishes.

Eqn(3a) is obtained by substituting in eqn(1) the following: 
 $$ {\mathcal F}^{m}_{halo}(r,\varphi,t) \: = \: -{\nu_{h}}^2 (r)z_{m}(r,\varphi,t) \eqno(4a)$$

\noindent where $\nu_{h}$ is the vertical frequency due to the dark matter halo in the unperturbed disk plane.
\begin{eqnarray}
\lefteqn{{{\mathcal F}^{m}_{self}(r,\varphi,t)} = -G{\int_{0}^{\infty}}{\Sigma(r^{\prime})r^{\prime} dr^{\prime}} } \nonumber \\
& & \times{ {\int_{0}^{2\pi}} \frac{[z_{m}(r,\varphi,t) -z_{m}(r^{\prime},\varphi^{\prime},t)]}{[r^2 + {r^{\prime}}^2 -2 r r^{\prime}\cos(\varphi -\varphi^{\prime}) + {z_0}^2]^{\frac{3}{2}}} d\varphi^{\prime} } \nonumber \: \: \: \: \: \: \: \: \: \: \: (4b) 
\end{eqnarray}
\noindent In the above equations $z_{0}$ is the softening parameter introduced to make the integrals regular at $r=r^{\prime}$. $z_{0}$ can be interpreted as a finite thickness of the disk.
The above integral eq.(3a) can be solved by recasting it into a matrix-eigenvalue problem. On a uniform grid with N radial points the resulting eigenvalue equation for the $m^{th}$ mode takes the form
$$ \left[{\omega_{m}}^2 -{2m{\Omega}(r_{i})}{\omega_{m}}\right] h_{m}(r_{i}) \:=\: \sum_{j=1}^{N} {\Lambda_{ij} h_{m}(r_{j})} \eqno(5a) $$

\noindent Where 
$$ \Lambda_{ij} \: = \: M_{ij} + \delta_{ij}{\left( {{\nu_{h}}^2}(r_{j}) +{{\nu_{dc}}^2}(r_{j}) - {m^2}{\Omega(r_{j})}^2 \right)} \eqno(5b) $$
$$ M_{ij} \: = \: {\Delta}r G \Sigma(r_{j})\left[ H(r_{i},r_{j}) - K_{m}(r_{i},r_{j})\right] r_{j} \eqno(5c) $$
And
$$ \nu_{dc}(r_{i}) \:=\: G{\int_{0}^{\infty}}{\Sigma(r^{\prime})H(r_{i},r^{\prime})r^{\prime} dr^{\prime}} \eqno(5d) $$

The above eq.(5a) can be rewritten in a more compact form:

$$ \left( {{\omega_{m}}^2}I + {\omega_{m}}R + \Lambda \right )h_{m} \: = \: 0 \eqno(6) $$
\noindent Where I, R and $\Lambda$ are the three N$\times$N real square matrices. $h_{m}$ is the eigenvector corresponding to the eigenvalue $\omega_{m}$.

The matrix elements are $I_{ij}=\delta_{ij}$, $R_{ij}=-2m\Omega(r_{j})\delta_{ij}$ and $\Lambda_{ij}$ as shown above in eq.(5b).


\subsection{ Asymmetric warp}

The bowl-shaped mode is an axisymmetric bending of the disk and the integral-sign warping mode is anti-symmetric in the 
azimuthal angle $(\varphi)$. In the linear regime each of the two bending modes behaves much like an independent oscillator in the system, so that 
we ignore any kind of energy cascading. This is shown to be valid in the last paragraph of sec. 3. By making a linear combination with proper time dependence of the two bending modes we are able to produce an asymmetric bending mode of 
the initially assumed axisymmetric disk. Thus according to eq.(2a) the asymmetric bending mode is described by

\begin{eqnarray}
{\mathcal Z}_{asym}(r,\varphi,t) & = & A_{\circ}z_{\circ}(r,\varphi,t) + A_{1}z_{1}(r,\varphi,t) \nonumber \\
                                & = & A_{\circ} \Re\{h_{\circ}(r)e^{i\omega_{\circ}t}\} +A_{1} \Re\{h_{1}(r)e^{i(\omega_{1}t - \varphi)}\} \nonumber \: \: \: \: (7) 
\end{eqnarray}

So that at any time $t=\tau$ we can analyse the behaviour of the 
resulting asymmetry.             
It was proved that the $m=0$ mode and the $m=1$ mode in a disk
described above are stable (see Hunter \& Toomre 1969), and
their stability was further confirmed by numerical work (see
SC88, Sparke 1995). Our numerical work also confirms the earlier result. 
Thus the resultant asymmetric bending mode can be assumed to be stable and 
it establishes a classic example of dynamical wave interference pattern in 
a gravitating medium.
The degree of asymmetry in the disk bending depends on the amplitude of the 
bowl-shaped mode which can be controlled by the free parameter $A_{\circ}$ 
in the problem. Thus by varying $A_{\circ}$ one can in principle produce 
a rich class of asymmetries that are seen in the observations. Recent 
observations show that asymmetric warps are indeed common and they exist in poor environments too (Garcia-Ruiz etal. 2002). This suggests that tidal interactions are not the only process that can produce an asymmetric warp. With our approach, we show how an asymmetric warp can be generated purely from the internal disk dynamics- assuming that these modes are generated due to some mechanism.

\subsection{Disk-Halo system}
We have adopted a truncated exponential disk with a $\cos^2$ tapering as introduced by SC88:

\begin{eqnarray*}
  \Sigma(r) & = & \Sigma_{0} \: e^{-r/R_{d}} \: \: \: \: \:   r\leq r_{trun} \\
            & = & \Sigma_{0} e^{-r/R_{d}}{\cos^2}\{ {\frac{\pi}{2}}{\frac{r-r_{trun}}{r_{out}-r_{trun}} }\} \: \: \: r_{trun}\leq r \leq r_{out} \\
            & = & 0  \: \: \: \: r \geq r_{out}  \: \: \: \: \: \: \: \: \: \: \: \: \: \: \: \: \: \: \: \: \: \: \: \: \: \: \: \: \: \: \: \: \: \: (8)
\end{eqnarray*}
 
\noindent where $R_d$ is the exponential disk scalelength and $\Sigma$ is the disk surface density.
The gradual truncation of the disk avoids spurious modes (SC88). In the calculations (Section 3), we have taken $r_{out}$ and 
$r_{trun}$ to be fixed at 6 and 5 disk scalelengths respectively.

The flattened dark matter halo is taken as a screened isothermal
halo which gives an asymptotically flat rotation curve (de Zeeuw \& Pfenniger 1988) :

$$  \rho(r,z) \: = \: \frac{\rho_{\circ}}{1 + {(r^2 + {\frac{z^2}{q^2}})}/{R_{c}^2}} \eqno (9) $$

\noindent where $R_{c}$ is the core radius and $q$ is the halo flattening parameter.     

The terminal velocity of dark matter halo is given by:

$$ v_t \:=\: 4\pi G\rho_{\circ}{R_c}^2 q \frac{sin^{-1}{\sqrt{1-q^2}}}{\sqrt{1-q^2}} \eqno (10)$$
 
\subsection{Numerical details}

Equation (3a) is a linear integral equation which is equivalent to an 
infinite dimensional eigenvalue equation (6). To simplify this
problem, a galactic disk of finite radius ($r_{out}$) is approximated as a 
system of N uniformly spaced rings. Then the resulting problem reduces 
to an N-dimensional eigenvalue problem in the eigenvalue $\omega_{m}$ 
and N-dimensional eigenvector $h_{m}$ describing the shape of the bent disk.
This requires the solution of an N$\times$N matrix.
The softening parameter $z_{\circ}$ is introduced to make the self-gravitating integral (eq.[4b]) regular along the diagonal line of the N-dimensional matrix.
\noindent We have carried out the numerical computation for different values of the matrix dimension N. We note that the eigen frequency of the ground state of either mode ($m=0$ \& $m=1$) is not very sensitive to the matrix dimension N or the softening parameter $z_{\circ}$ in the problem. Numerical results show that as the value of N increases or the softening  $z_{\circ}$ decreases, the ground-state eigen-frequency $\omega_1$ of the warping mode ($m=1$) reduces extremely slowly. For example over an augmentation of 200 rings the change in $\omega_1 \sim 5.0$\%. For convenience we have taken N=80 for our calculations. A similar calculation was done for the case of the bowl-shaped mode ($m=0$) but it did not show any definite trend unlike the $m=1$ case. But the change in the frequency($\omega_{\circ}$) occurs only in the third decimal place and its quite arbitrary. The insensitivy of the ground-state mode frequencies ($\omega_{\circ}$ \& $\omega_1$) on the softening is obvious because these are the integral properties of the disk.

\noindent However the ground-state mode shape, $h_m$, in either case is quasi-sensitive to the softening parameter. A smooth bending of the disk was obtained for typical values of the softening $z_{\circ}$ $\sim$ inter ring spacing. We note that a large value of the softening reduces the value of the self-gravitating integral, hence it becomes inefficient to make all the rings precess in a synchronized manner, thus any resultant warping is dissolved. On the other hand if $z_{\circ}=0$ it is obvious that the self-gravitating integral diverges numerically. Since we are not interested in investigating the structure of the either mode in the resolution $\sim$ inter-ring separation, we keep the softening parameter $z_{\circ}$ $\sim$ inter-ring spacing and this gives us a satisfactory result for the mode shape. The sensitivity of the mode shape on the softening is understandable because a global mode shape depends on the local details of the disk.

For $m=0$,  i.e. bowl-shaped mode, the eigenvalue equation (6) is linear in 
${\omega_{\circ}}^2$. But for $m=1$, i.e. a warping mode, the problem 
reduces to a quadratic eigenvalue problem (known as QEP in the literature) 
in $\omega_{1}$.
The standard way to solve the QEP (6) for the $m$=1 mode is to reduce it to a generalized eigenvalue problem (GEP) of the form $Ax=\omega_{m}Bx$ of twice the matrix dimension 2N see Bai et al. (2000). This GEP is commonly called a linearization of the QEP (6). We solve the linearized QEP numerically by using standard technique for diagonalization. For further details about solving QEP the readers are referred to Guo, C. -H (2004) and Higham \& Kim, (2000). 

\section {Results}
Recent observations have shown that the galactic disks bend with a 
variety of shapes starting from symmetric 'S' shaped warps to asymmetric warps and also 'L' shaped 
and 'U' shaped warps (Section 1).
We next show that this rich variety of warp shapes observed can be explained naturally right from the internal disk dynamics by a superposition of the $m=1$ and $m=0$ modes.
The superposition picture proposed and studied here is verified for a few cases in the results from the N-body simulations, and there is evidence for an
asymmetric warp arising due to a superposition of m=1 and m=0, or m=1
and m=2 modes (Revaz \& Pfenniger 2005, personal communication), this
requires further study using N-body simulations.  
Since we superpose two modes, the range of values spanned cover the two 
frequencies $\omega_0 $, $\omega_1$, the superposition amplitudes 
 $A_0$, $A_1$, and the two phases. The unmodulated amplitudes $h_1$ and $h_0$ 
(see eq. [2b]) are free upto a linear multiplication factor.
These and the frequencies are obtained from our modal analysis.
The ratios of $\omega_0$/$\omega_1$ are seen to vary from 8-20 when the flattening of the dark matter halo (q) varies from 0.5 to 0.9.
(see the discussion at the end of this section).
We do not have any apriori physical basis to choose the values of
$A_0$ and $A_1$, these are set by the generation process of the modes-
hence we cover a reasonable range for these.

\noindent All the numerical calculations are done in units of G=$M_d$=$R_d$=1 where $M_d$ is the disk mass. In order to show the basic results first, we have fixed the dark matter halo parameters at $R_c =2.0$ ( in units of $ R_{d}$) , $\rho_{\circ}=0.0117$ ( in units of $M_d/R_{d}^3$) with a halo flattening parameter $q=0.7$. These give rise to the terminal velocity of the halo as $v_{t} = 0.68$ ( in units of $\sqrt{GM_d/R_d})$.

\noindent In order to check our numerical calculations with the previous works by SC88 and Sparke, 1995 we have reproduced the ground state mode shape for the modes $m$=0 and $m$=1 for some typical disk and halo parameters as can be seen in Figure 1. The amplitude of these modes are measured with respect to the inner disk plane and these we call as the unmodulated amplitudes. In addition to this, Figure 1 also shows the sensitivity of the amplitudes of these modes with respect to disk edge. To study the sensitivity we have varied $r_{trun}$ keeping the disk outer edge fixed. The solid, dashed and dotted lines are for $r_{trun}$=5.0, 5.5 and 5.95 in both the figure panel. The disk density starts deviating from the exponential from $r=r_{trun}$ and smoothly tapers to zero at $r=r_{out}$. The dotted curves show the behaviour of the modes when the disk is almost abruptly truncated and in this case too the two modes m=0 and m=1 behave in the same way. Also note that the amplitudes of m=0 modes are slightly less than that of m=1 modes.
The eigen frequencies corresponding to these ground states are not quite sensitive to the behaviour of disk edge which can be seen from the SC formula for the modified tilt mode(eqn(21) in SC88). This actually confirms what we get numerically for the m=1 mode. However for the m=0 modes the eigen frequencies vary from 0.1315 to 0.1296 as $r_{trun}$ varies from 5.0 to 5.95.

\noindent In Figure 2 we have shown four subplots of an asymmetric warp generated with various values of the controlling parameter $\zeta = A_{\circ}/A_{1}$ which we will use as an indicator of the relative strength of the bowl-shaped mode. $A_{\circ}=1.0$ and $A_1 =1.0$ connotes that the bowl-shaped mode and the 'S' shaped warping mode both are present with their full unmodulated amplitudes, $h_m$ (see eq. [2b]). 
These plots are made for an epoch of $\tau_{\circ}/8$ as an illustrative 
case, where  $\tau_{\circ} = {2{\pi}}/\omega_{\circ}$ is the
 characteristic time scale in the problem
and $\omega_{\circ}$
 is the ground state eigenfrequency of the bowl-shaped mode. 
As we move from Fig. 2a to Fig. 2d, 
the value of the controlling parameter $\zeta$ 
increases, that is, the relative strength of the bowl-shaped 
mode  increases. Since in our analysis bowl-shaped mode is the 
only cause of asymmetry in the bending, we can see from Fig. 2d, that
the warp is most asymmetric i.e. the degree of asymmetry is the highest.       

\noindent In Figure 3, we have produced asymmetric warps at
different epochs ($t$) keeping the controlling parameter $\zeta
=1.0$. The values of the sampling times $t$ were so chosen that
as we move from Fig. 3a  to Fig. 3d, we see the diverse
phenomena of disk warping including symmetric as well as
asymmetric ones. This is due to the characteristic oscillation
of the disk in a bowl-shaped mode. The disk oscillates from
'cupped upward' to flat, to 'cupped downward' and back again with 
its typical ground state frequency $\omega_{\circ}$. During this 
flexing of the disk it gets a chance to interfere with the S-shaped mode
 of the disk and this produces a rich class of dynamical asymmetric figures. 
In a real galaxy the bowl-shaped mode may decay due dissipative effects. 
Even then the mild asymmetries in disk warping which are seen 
in observations can be considered as a signature of the bowl-shaped mode.

\noindent In the above two figures (Figs. 2-3) we have shown 
the warping of the disk with respect to the inner unwarped disk plane 
because in actual observations the warping is quantified 
based on measurements with respect to the inner disk plane. 
Next, we define a quantitative measure of the asymmetry seen in a disk galaxy,
the asymmetry index, as follows (see also Sanchez-Saavedra et al. (2003)):

\begin{eqnarray*}
\alpha_{asym} =\frac{|\alpha_{right} -\alpha_{left}|}{\alpha_{right} +\alpha_{left}} \: \: \: \: \: \: \: \: \: \: \: \: \: {if\: \: \: \: {\alpha_{right}^2 + \alpha_{left}^2\neq 0}} \: \: \: \: \: \: \: \: \: \: \: \: \: \: \: (11) 
\end{eqnarray*}

\noindent where $\alpha_{right}$ is the angle between the line joing the 
centre to the outer most point on the right hand side of the particular 
warp and the inner disk plane. $ \alpha_{left}$ is defined similarly. 
Thus $\alpha_{asym}$ is the normalized or fractional value of the asymmetry. 
In terms of this definition, the results in Figures 2 and 3
give an asymmetry of $\alpha_{asym} \sim 0.2-0.4$. Thus our
results for asymmetry agree with the typical range of measured
values in Sanchez-Saavedra et al. (2003) and the asymmetries when measured from the figures given in appendices (A \& B) in Schwarzkopf \& Dettmar (2001).
For example the value of asymmetry index for the S-shaped asymmetric warp(galaxy name- AM 1134-323(R)) when measured from Fig(6) in Schwarzkopf \& Dettmar (2001) gives a value 0.33. This agrees closely with what we get from Fig(2). Of course true comparisons with observations is a bit tricky because of the free parameters($A_0$ and $A_1$) involved in our problem. Our basic idea is to show that with this modal superposition approach, one can reproduce various asymmetric warps that are seen in observations.

Clearly when $\alpha_{right}=\alpha_{left}$ the asymmetric index $\alpha_{asym}$ vanishes. This implies a purely symmetric warping of the disk. Note that eq.(11) avoids the possibility of $\alpha_{right}$ and $\alpha_{left}$ being simultaneosly zero as the case is not physically meaningful.  
When $\alpha_{right}=0$ or $\alpha_{left}=0$ we obtain unity value of the 
asymmetric index. So $\alpha_{asym} = 1$ (maximum asymmetry) denotes a 
perfectly 'L' shaped warping of the disk. This can be called a one-sided warp.
The values of $ \alpha_{asym}$ vary from 0.0 to 1.0. It is obvious that most of the geometrical shapes of galactic warps seen in observations can be classified using a single parameter $\alpha_{asym}$. Jimenez-Vicente et al. (1997) used a different set of 3 parameters to describe the shape of warps.  However neither their model nor $\alpha_{asym}$ alone can describe the peculiar shape of the warp of our Galaxy. But $\alpha_{asym}$ being a single parameter is very useful in depicting the underlying asymmetries in warps.
 Given the extreme importance of this parameter, we next study 
the dependence of $\alpha_{asym}$ on the dark matter halo parameters.

\noindent Figure 4 shows the variation of the asymmetry index, $\alpha_{asym}$,
 due to the variation in $q$, the flattening parameter of the dark matter halo
  for a fixed value 
 of the controlling parameter ($\zeta$=1) and at an epoch $t=\tau_{\circ}/8$ 
 as a special case. 
As the flattening parameter of a halo of fixed mass is varied its central 
density, $\rho_{\circ}$, and the core radius, $R_c$ are bound to change. 
Therefore we need to calculate the $\rho_{\circ}$ and $R_c$ as
functions of $q$. By imposing two constraints: the mass within a
thin spheroidal shell(Binney \& Tremaine 1987, pg 54) and the
terminal velocity ($v_t$,  see eq. [10]) of the halo should be independent of $q$,
 we obtain $\rho_{\circ}(q)$ and $R_c(q)$
 in terms of their spherical counterparts as (see Narayan et al. 2005):

$$ \rho_{\circ}(q) \:=\:\rho_{\circ}(1){\frac{1}{q}}{\left( \frac{e}{sin^{-1}e}\right)}^3, \: \: \: \: \: 
     R_{c}(q) \:=\: R_{c}(1){\left( \frac{sin^{-1}e}{e}\right)} \eqno (12)   $$ 
 
\noindent where $e=  (1-q^2)^{1/2} $. As the halo  becomes more flattened 
(smaller $q$), the resulting asymmetry $\alpha_{asym}$ goes down which 
means that the warp becomes
more symmetric. Conversely, as the halo becomes more spherical the 
asymmetric index $\alpha_{asym}$ goes up showing that asymmetric warps 
are more likely to be found in a less oblate dark matter halo. This can be explained as follows.
As a constant-mass halo is flattened, the ground state
eigenfrequency ($\omega_{\circ}$) of the bowl-shaped mode increases, 
 so that the disk flexes up and down more rapidly. 
 Therefore the characteristic timescale ($\tau_{\circ}$) of the 
 mode decreases and thereby its unmodulated amplitude reduces faster at the
measured epoch, this in turn reduces the asymmetry of the warps.
 While this figure illustrates an important
physical point from the model, its direct observational
verification is not possible.
 
\noindent In Figure 5 we plot the variation of the asymmetry index, 
$\alpha_{asym}$, with the core radius $R_c$ of the dark matter halo 
(given in units of $R_d$), again for a fixed value of the controlling 
parameter ($\zeta$=1) and at an epoch $t=\tau_{\circ}/8$ and for a halo 
flattening of $q=0.75$. In contrast to Fig. 4, here as $R_c$ increases, the halo mass within a
given radius goes up. Fig. 5 shows that asymmetry in warps are likely to more in a halo with a smaller core radius; while as $R_c$ increases, $\alpha_{asym}$ 
goes down. This is because as $R_c$ and hence the halo mass increases,  
  the frequency $\omega_{\circ}$ of the bowl-shaped mode is raised,
and this results in a smaller intrinsic or unmodulated amplitude of 
the bowl-shaped mode.
So in absence of an appreciable amplitude of the bowl-shaped mode, 
the resulting asymmetry in the disk bending also reduces. 
This result of decreasing asymmetry in warps for higher mass galaxies
is in agreement with observations (Castro-Rodriguez et al. 2002), which 
further supports our model.

\noindent The variation of the asymmetry index $\alpha_{asym}$ on the halo mass can be seen from the Fig. 5. It is explained above that as $R_c/R_d$ increases and hence the halo mass increases the asymmetry index goes down. While as the halo mass decreases the amplitude of warping mode ($m=1$) starts decreasing and in the limiting case when the dark matter halo is absent $m=1$ mode turns out to be a trivial tilting of the disk. And it is hard to generate $m=0$ mode in the absence of a dark matter halo. Thus in the limiting case when halo mass is tending to zero, the asymmetry index starts falling down to zero and however as we have checked that this happens when $R_c/R_d$ starts becoming a fraction which is not a physically acceptable regime of $R_c/R_d$ for any reasonable disk-halo system known.

So far we have considered a screened isothermal ($\rho \propto r^{-2}$ at large radii) dark matter halo, producing a flat rotation curve, for the generation of asymmetric warps. The flaring of HI in the outer region of our Galaxy favours a dark matter halo with a steeply falling density profile ($\rho \propto r^{-3}$ or $r^{-4}$ at large radii) ( Narayan et al. 2005). Note that $\rho \propto r^{-3}$, at large radii, is the popular NFW density profile (Navarro et al. 1996) for the dark matter halo. We have checked that the amplitudes of the two modes ($m=0$ \& $m=1$) and their ground state eigen frequencies behave in a similar fashion in both cases when $\rho \propto r^{-2}$  and $\rho \propto r^{-3}$ or $r^{-4}$ at large radii. Hence we can say that qualitative behaviour of the asymmetry would remain similar. However we would like to carry a detailed investigation of the asymmetric warp with respect to differnet halo profiles in a future work.      

In Figure 6 we show one of the extreme form of asymmetry of warp shapes, namely an L-shaped mode (see above discussion following eq.[11]).
In principle this can be achieved by reinforcing the net amplitude of 
bending resulting from the superposition of the two modes to zero on one 
side of the disk. 
 We get an approximately L-shaped mode 
 for $A_0=0.55$ and $A_1=0.20$ for the usual input values of $q$=0.76 and 
 $R_c$=2.0. Clearly the root of such an extreme form of asymmetric warp 
 lies in the fine balancing between the two modes in the process of 
 superposition.
  Nearly 5 \% of the asymmetric warps observed have this structure (Sanchez-Saavedra et al. 2003), which can be explained by our approach.

Another form of deviation from an S-shaped distribution is a U-shaped warp, which comes
about naturally when the amplitude $A_0$ of the bowl-shaped ($m=0$) mode is 
enhanced w.r.t. the amplitude $A_1$ of the warping ($m=1$) mode (see 
Figure 7). Nearly 7 \%   of the asymmetric warps observed have this 
structure  (Sanchez-Saavedra et al. 2003). 
A little juggling of $A_1$ and $A_0$ values is required to obtain the
L-shaped and the U-shaped modes, unlike the case of the other
asymmetric warps. This perhaps explains their lower observed frequency.

Thus our approach can naturally produce the variety of asymmetry seen in 
disk galaxies, including the peculiar L-shaped and U-shaped warps. 
The resulting values of asymmetry (Figs. 4-5) agree with the
range of observed values (Sanchez-Saavedra et al. 2003 and Schwarzkopf \& Dettmar 2001).

In order to compare the resulting asymmetry with the observed values, 
it is useful to obtain the true value of $\tau_0$. We have checked that 
this lies in the range of $\sim 0.5 - 1.0$ Gyr and $\tau_1$ 
lies in the range of $\sim 6 - 12$ Gyr, for the realistic galaxy models 
with a range of $R_d \sim 2-3$ kpc, $M_D$, the disk mass$ \sim 2
- 6\times 10^{10} M_{\odot}$,
 $ R_c \sim 2 R_d $, $q$ = 0.7 and the ratio of the halo mass to the 
disk mass is$\sim$ 1.4 within the outer radius ($\sim 6 R_d$).
The ratio of the frequencies $\omega_0 / \omega_1$ is $\sim
10$, note that this value is obtained for q=0.7. For other values of q the ratio
$\omega_0 / \omega_1$  varies. Numerical calculations yield a range of values varying from 8-20 for q varying from 0.5 to 0.9 respectively.
In the absence of any dissipation, the superposition of the two modes will 
show a cyclical variation and hence the above values represent the typical 
asymmetries seen. Even if there is dissipation, 
so long as there is a regeneration of the modes 
(see Section 1), then such asymmetric warps can 
recur and hence can be seen effectively over much larger timescales.
We note that since the natural frequencies of the two modes are
so different, we are justified in treating the problem as a simple
linear superposition  rather than having to take account of the
mode coupling.

\section{Conclusions}

We propose and study the origin of the asymmetric warps in spiral galaxies 
as arising due to the superposition of the standard S-shaped warps 
($m=1$ mode) and a bowl-shaped warp ($m=0$ mode). 
This is a simple but general model for generating
asymmetric warps that is independent of whether these modes arise
due to tidal interaction, or gas accretion, or bending instabilities, or any other perturber.
We do a modal analysis of a disk embedded in a dark matter halo, and obtain 
the solutions for these two modes which are then linearly superposed. 
In the characteristic oscillation of the $m=0$ mode, the disk
oscillates from `cupped upwards' to flat, to `cupped downwards'
and back again. During this flexing of the disk, it gets a
chance to interefere with the S-shaped ($m=1$) mode of the disk,
and this produces a rich class of dynamical, asymmetric warps.
 The results 
obtained naturally explain the wide variety of asymmetric 
warps observed in spiral galaxies, including the peculiar L-shaped 
and U-shaped warps.

\noindent {\bf Acknowledgements:}

We are happy to acknowledge useful e-mail correspondence on the $m=0$ mode with Linda Sparke. We are grateful to Yves Revaz and Daniel Pfenniger for analyzing 
some of their N-body simulations data to check that there is evidence for the
superposition idea proposed here. We also thank them and Francoise Combes, and  Martin Lopez-Corredoira for critical comments on the manuscript. We thank the referee Eduardo Battaner for constructive comments on the paper. The numerical package LAPACK (see www.netlib.org) was used for solving the matrix diagonalization. K.S. thanks the CSIR-UGC, India for a Senior research fellowship.

\newpage

\section {References}

\noindent  Bai, Z., Demmel, J., Dongarra, J., Ruhe, A. and van der Vorst, H. 2000, editors. Templates for the Solution of Algebraic Eigenvalue Problems: A Practical Guide (Philadelphia: SIAM).

\noindent Battaner, E., Florido, E., \& Sanchez-Saavedra, M. L. 1990, A \& A, 236, 1 

\noindent Binney, J., Jiang, I.-G., \& Dutta, S. 1998, MNRAS, 297, 1237

\noindent Binney, J. \& Tremaine, S. 1987, Galactic Dynamics
(Princeton: Princeton Univ. Press)

\noindent Bosma, A. 1978, Ph.D. thesis, University of Groningen

\noindent Burton, W.B. 1988, in Galactic and Extragalactic Radio Astronomy, ed. G.L. Verschuur \& K.I. Kellermann (Berlin: Springer), 295

\noindent Castro-Rodriguez, N., Lopez-Corredoira, M.,
Sanchez-Saavedra, M.L., \& Battaner, E. 2002, A\&A, 391, 519

\noindent de Zeeuw, T. \& Pfenniger, D. 1988, MNRAS, 235, 949

\noindent Dubinski, J., \& Kuijken, K. 1995, ApJ, 442, 492

\noindent Garcia-Ruiz, I., Sancisi,R., \& Kuijken,K. 2002, A\&A, 394, 769

\noindent Guo, C. -H., 2004, Linear Algebra Appl. , 385, 391

\noindent Higham, N.J., \& Kim, H. -M., 2000, IMA J. Numer. Anal., 20, 499

\noindent Hunter,C., \& Toomre, A. 1969, ApJ, 155,747

\noindent Jimenez-Vicente, J., Porcel, C., Sanchez-Saavedra, M.L., \& Battaner, E. 1997, Ap\&SS, 253, 225  

\noindent Lopez-Corredoira, M., Betancort-Rijo, J., \& Beckman, J.E.
   2002, A\&A, 386, 169

\noindent Narayan, C. A., Saha, K., \& Jog, C. J. 2005, A \& A, 440, 523 

\noindent Navarro, J.F., Frenk, C.S., \& White, S.D.M. 1996, ApJ, 462, 563

\noindent Nelson, R.W., \& Tremaine, S. 1995, MNRAS, 275, 897

\noindent Reshetnikov, V., \& Combes, F. 1998, A \& A, 337, 9
 
\noindent Reshetnikov, V., \& Combes, F. 1999, A \& A S, 138, 101

\noindent Revaz, Y. \& Pfenniger, D. 2004, A \& A, 425, 67

\noindent Sanchez-Saavedra, M.L., Battaner, E., \& Florido, E. 
  1990, MNRAS, 246, 458

\noindent Sanchez-Saavedra, M.L., Battaner, E., Guijarro,A., 
 Lopez-Corredoira, M.,  \& Castro-Rodriguez, N. 2003, A \& A, 399, 457 

\noindent Sancisi, R. 1976, A \& A, 53, 159

\noindent Schwarzkopf, U., \& Dettmar, R.-J. 2001, A \& A, 373, 402

\noindent Sellwood, J. 1996, ApJ, 473, 733

\noindent Semelin, B. \& Combes, F. 2005, A \& A, in press, (astro-ph/0506589)

\noindent Sparke, L.S. 1995, ApJ, 439, 42

\noindent Sparke, L.S., \& Casertano, S. 1988, MNRAS, 234, 873 (SC88)

\noindent van der Kruit, P., \& Searle, L. 1981, A \& A, 95, 105

\newpage

\bigskip
\begin{figure}

{\rotatebox{270}{\resizebox{8cm}{8cm}{\includegraphics{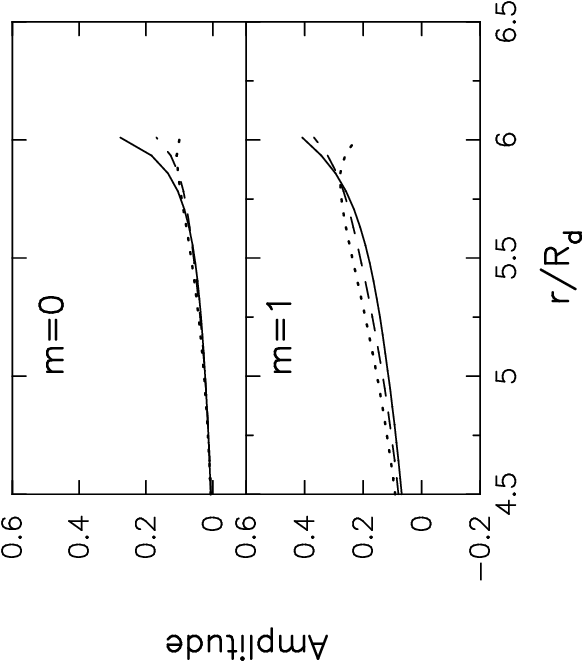}}}}
\bigskip

\noindent{\bf Figure 1.}
The ground state mode shapes and their sensitivity w.r.t. disk edge for the two modes $m$=0 and $m$=1. The mode amplitudes are in arbitrary units. The solid, dashed and dotted lines are for $r_{trun}$=5.0, 5.5 and 5.95 respectively and $r_{out}$=6.0. Here, the dark matter halo flattening $q = 0.7$ and  the core radius $R_c = 2.0$. 
\end{figure}

\begin{figure}
{\rotatebox{270}{\resizebox{8.5cm}{8.5cm}{\includegraphics{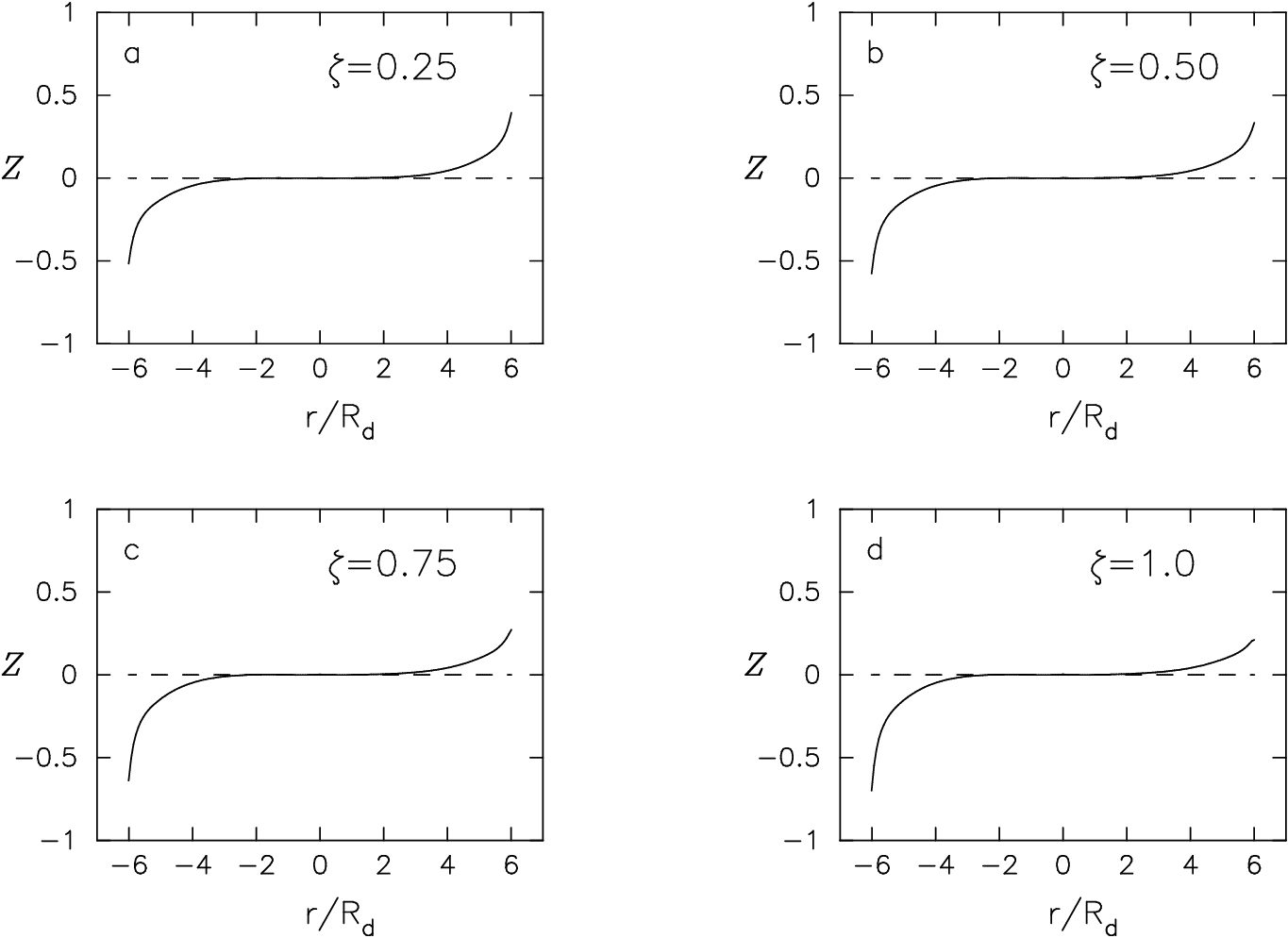}}}}
\bigskip

\noindent{\bf Figure 2.}
A plot of the bending amplitude of asymmetric warp versus radius, for various
values of the parameter $\zeta = A_0/A_1$ denoting the ratio of the superposition amplitudes, at an epoch $t=\tau_{\circ}/8$. As $\zeta$ increases to 1, the asymmetry becomes prominent. Here, the dark matter halo flattening $q = 0.7$ and  the core radius $R_c = 2.0$. 
\end{figure}

\begin{figure}
{\rotatebox{270}{\resizebox{8.5cm}{8.5cm}{\includegraphics{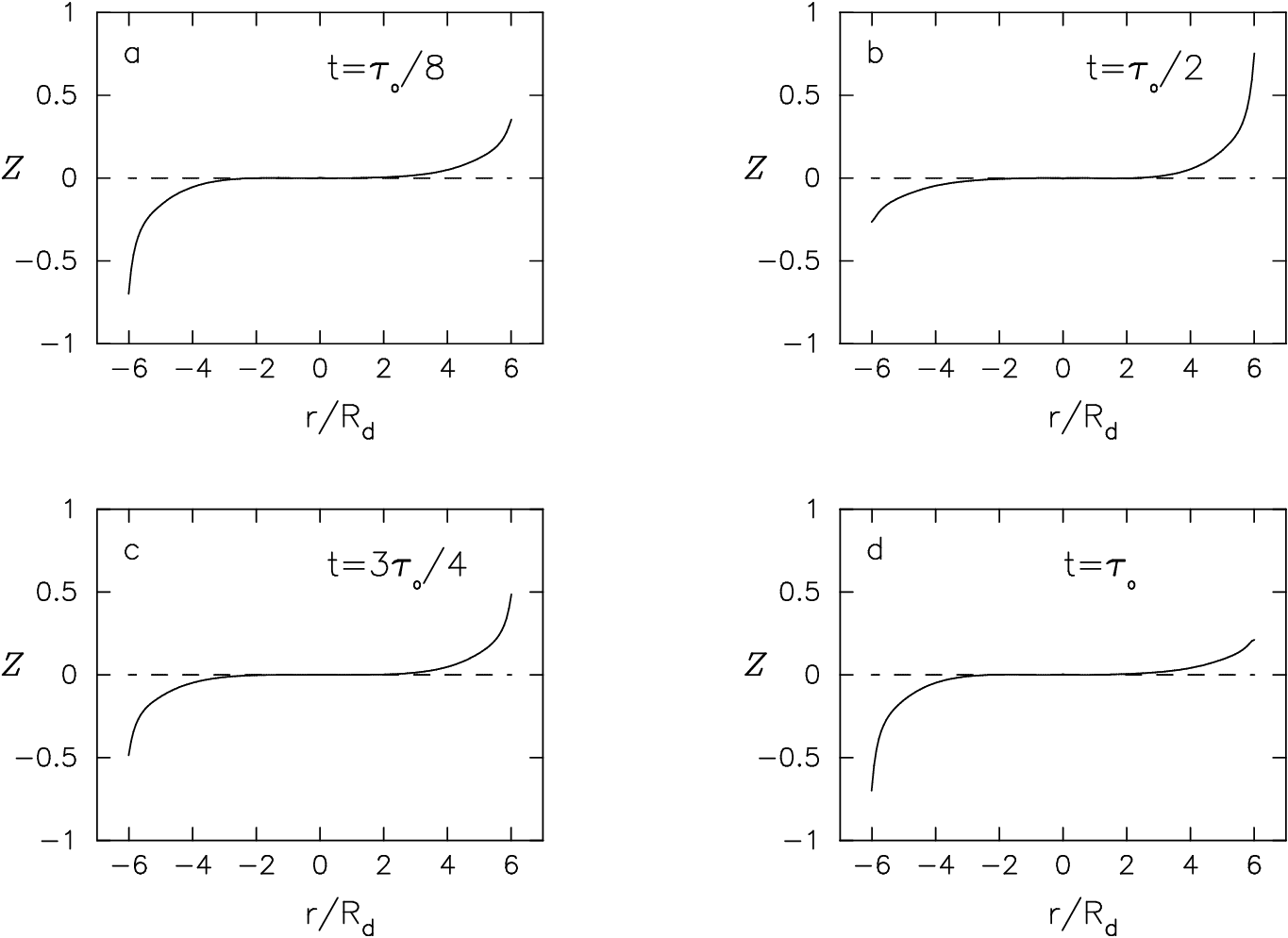}}}}
\bigskip

\noindent {\bf Figure 3.}
A plot of the bending amplitude of the asymmetric warp versus radius, at different epochs $t$. As $t$ goes through the cycle of the bowl-shaped mode it generates various kinds of disk warping from an asymmetric to a symmetric and then an asymmetric case. Again, the dark matter halo flattening $q = 0.7$ and the core radius $R_c = 2.0$. 
\end{figure}

\begin{figure}
{\rotatebox{270}{\resizebox{8cm}{8cm}{\includegraphics{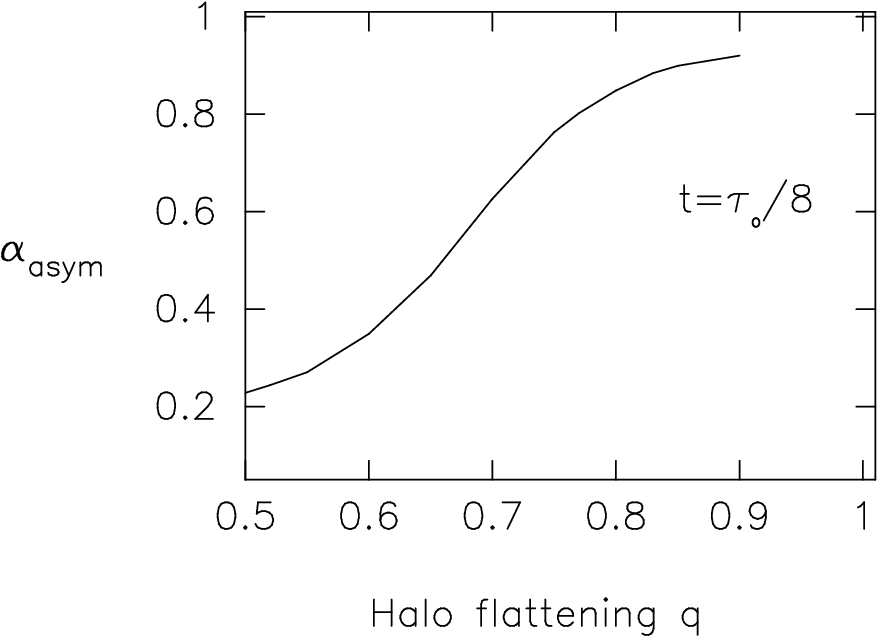}}}}
\bigskip

\noindent {\bf Figure 4.}
A plot of the asymmetry index, $\alpha_{asym}$, versus $q$, the dark matter halo flattening. As the halo becomes more spherical, the asymmetry in warps goes up. Here, the dark matter halo is flattened in such a way as to conserve the ratio of halo mass to disk mass within optical radius of the disk, and the terminal velocity of the halo. 
\end{figure}

\begin{figure}
{\rotatebox{270}{\resizebox{8cm}{8cm}{\includegraphics{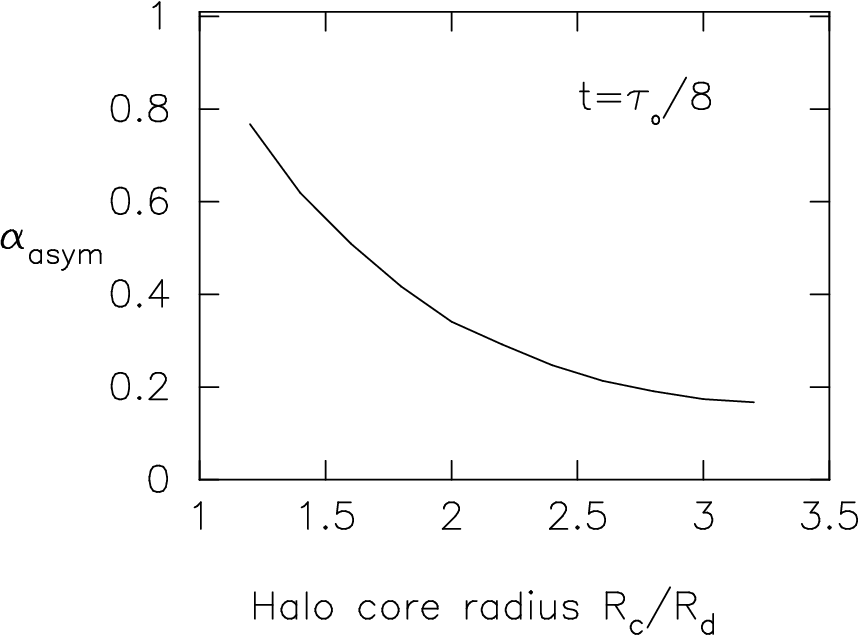}}}}
\bigskip

\noindent {\bf Figure 5.} A plot of the asymmetry index, $\alpha_{asym}$, versus $R_c$, the dark matter core radius. This shows that as the halo core radius and the halo mass increase, the asymmetry in warps reduces. Here, the dark matter halo flattening is kept constant at $q=0.75$.  
\end{figure}

\begin{figure}
{\rotatebox{270}{\resizebox{8cm}{8cm}{\includegraphics{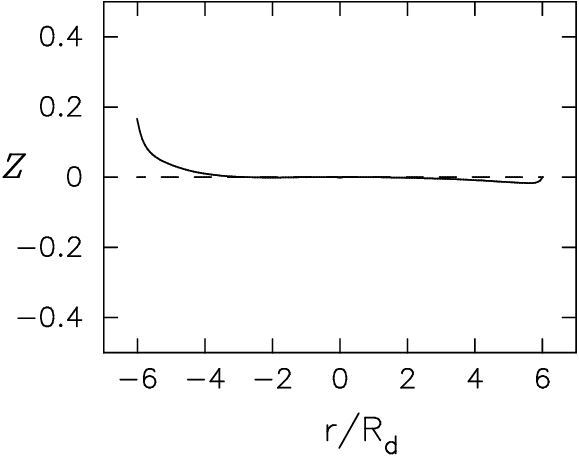}}}}
\bigskip

\noindent {\bf Figure 6.}
A plot of bending amplitude versus radius which shows an L-shaped asymmetric warp produced, by setting $A_0=0.55$ and $A_1=0.20$ at an epoch $t=\tau_{\circ}/8$. The dark matter halo flattening is kept at $q=0.76$ and $R_c=2.0$. 
\end{figure}

\begin{figure}
{\rotatebox{270}{\resizebox{8cm}{8cm}{\includegraphics{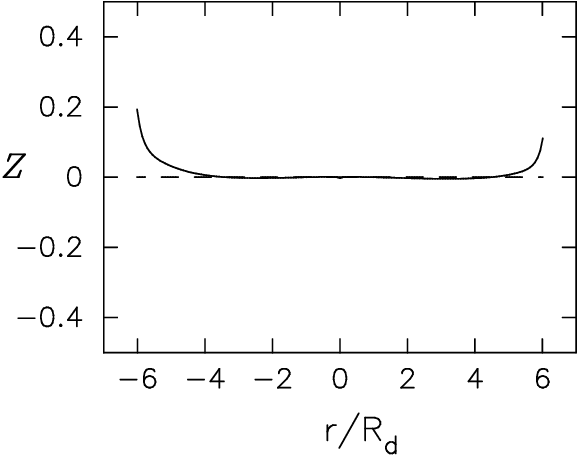}}}}
\bigskip

\noindent {\bf Figure 7.}
A plot of bending amplitude versus radius which shows an U-shaped asymmetric warp produced by setting $A_0=1.0$ and$A_1=0.1$ at an epoch $t=\tau_{\circ}/8$. The dark matter halo flattening is kept at $q=0.76$ and $R_c=2.0$. 
\end{figure}

\end{document}